\documentclass{article}
\usepackage{multicol}
\usepackage{amssymb,amsmath}

\advance \topmargin by -\headheight
\advance \topmargin by -\headsep
     
\evensidemargin \oddsidemargin
\advance\hoffset by -3mm  
\advance\voffset by  8mm  
     
\newcommand\rf[1]{(\ref{eq:#1})}
\newcommand\lab[1]{\label{eq:#1}}
\newcommand\nonu{\nonumber}
\newcommand\br{\begin{eqnarray}}
\newcommand\er{\end{eqnarray}}
\newcommand\be{\begin{equation}}
\newcommand\ee{\end{equation}}

\newcommand\foot[1]{\footnotemark\footnotetext{#1}}
\newcommand\lb{\lbrack}
\newcommand\rb{\rbrack}

\newcommand\llb{\left\lbrack}
\newcommand\rrb{\right\rbrack}

\newcommand\lcurl{\left\{}
\newcommand\rcurl{\right\}}
\renewcommand\({\left(}
\renewcommand\){\right)}
\newcommand\bv{\bigm\vert}               
\newcommand\bgv{\bigg\vert}              

\newcommand\bc{\begin{center}}
\newcommand\ec{\end{center}}

\newcommand\partder[2]{\frac{{\partial {#1}}}{{\partial {#2}}}}

\renewcommand\b{\beta}

\renewcommand\d{\delta}

\newcommand\eps{\epsilon}
\newcommand\vareps{\varepsilon}
\newcommand\g{\gamma}
\newcommand\G{\Gamma}

\newcommand\h{\frac{1}{2}}
\renewcommand\k{\kappa}
\renewcommand\l{\lambda}
\renewcommand\L{\Lambda}
\newcommand\m{\mu}
\newcommand\n{\nu}

\newcommand\vp{\varphi}
\renewcommand\P{\Phi}
\newcommand\pa{\partial}

\newcommand\pr{\prime}

\renewcommand\r{\rho}
\newcommand\s{\sigma}
\renewcommand\S{\Sigma}
\renewcommand\t{\tau}
\renewcommand\th{\theta}

\newcommand\wti{\widetilde}

\newcommand\cA{{\mathcal A}}

\newcommand\cE{{\mathcal E}}
\newcommand\cF{{\mathcal F}}

\newcommand\cJ{{\mathcal J}}

\newcommand\cV{{\mathcal V}}

\newcommand{\ct}[1]{\cite{#1}}
\newcommand{\bib}[1]{\bibitem{#1}}
\newcommand\PRL[3]{\textsl{Phys. Rev. Lett.} \textbf{#1}, #3 (#2)}
\newcommand\NPB[3]{\textsl{Nucl. Phys.} \textbf{B#1}, #3 (#2)}

\newcommand\PRD[3]{\textsl{Phys. Rev.} \textbf{D#1}, #3 (#2)}

\newcommand\PLB[3]{\textsl{Phys. Lett.} \textbf{#1B}, #3 (#2)}
\newcommand\CQG[3]{\textsl{Class. Quantum Grav.} \textbf{#1}, #3 (#2)}

\newcommand\AoP[3]{\textsl{Ann. of Phys.} \textbf{#1}, #3 (#2)}

\newcommand\IJMPA[3]{\textsl{Int. J. Mod. Phys.} \textbf{A#1}, #3 (#2)}

\newcommand\Xdot{\stackrel{.}{X}}
\newcommand\xdot{\stackrel{.}{x}}

\newcommand\etadot{\stackrel{.}{\eta}}

\begin{document}

\title{Space-Time Compactification Induced By Lightlike Branes}
\author{Eduardo Guendelman$^1$, Alexander Kaganovich$^1$, 
Emil Nissimov$^2$ and Svetlana Pacheva$^2$\\
\small\it $(^1)$ Department of Physics, Ben-Gurion University of the Negev,
Beer-Sheva, Israel \\[-1.mm]
\small\it $(^2)$ Institute for Nuclear Research and Nuclear Energy,
Bulgarian Academy of Sciences, Sofia, Bulgaria  \\[-1.mm]
\small\it email: guendel@bgu.ac.il, alexk@bgu.ac.il,
nissimov@inrne.bas.bg, svetlana@inrne.bas.bg}
\date{ }
\maketitle

\begin{abstract}
The aim of the present paper is two-fold. First we describe the Lagrangian
dynamics of a recently proposed new class of {\em lightlike} $p$-branes and
their interactions with bulk space-time gravity and electromagnetism in a
self-consistent manner. Next, we discuss the role of {\em lightlike} branes
as natural candidates for {\em wormhole} ``throats'' and exemplify the
latter by presenting an explicit construction of a new type of asymmetric
wormhole solution where the {\em lightlike} brane connects a ``right''
universe with Reissner-Nordstr{\"o}m geometry to a ``left''
Bertotti-Robinson universe with two compactified space dimensions.
\end{abstract}




\begin{multicols}{2}
\section{Introduction}

Lightlike branes (\textsl{LL-branes} for short) play an increasingly significant role 
in general relativity and modern non-perturbative string theory.
Mathematically they represent singular null hypersurfaces in Riemannian
space-time which provide dynamical description of various physically important 
cosmological and astrophysical phenomena such as:

(i) Impulsive lightlike signals arising in cataclysmic astrophysical events 
(supernovae, neutron star collisions) \ct{barrabes-hogan}; 

(ii) Dynamics of horizons in black hole physics -- the so called ``membrane paradigm''
\ct{membrane-paradigm};

(iii) The thin-wall approach to domain walls coupled to 
gravity \ct{Israel-66,Barrabes-Israel,Dray-Hooft}.

More recently, the relevance of \textsl{LL-branes} in the context of
non-perturbative string theory has also been recognized, specifically, as the so called
$H$-branes describing quantum horizons (black hole and cosmological)
\ct{kogan-01}, as Penrose limits of baryonic $D$-branes
\ct{mateos-02}, \textsl{etc} (see also Refs.\ct{nonperturb-string}).

A characteristic feature of the formalism for \textsl{LL-branes} in the
pioneering papers \ct{Israel-66,Barrabes-Israel,Dray-Hooft}
in the context of gravity and cosmology is that they have been exclusively
treated in a ``phenomenological'' manner, \textsl{i.e.}, without specifying
an underlying Lagrangian dynamics from which they may originate. As a
partial exception, in a more recent paper \ct{barrabes-israel-05} brane actions in 
terms of their pertinent extrinsic geometry
have been proposed which generically describe non-lightlike branes, whereas the 
lightlike branes are treated as a limiting case. 

On the other hand, in the last few years we have proposed in a series of papers 
\ct{LL-brane-main,inflation-all,our-WH,rot-WH} a new class of concise
manifestly reparametrization invariant world-volume Lagrangian actions, 
providing a derivation from first principles of the \textsl{LL-brane} dynamics.
The following characteristic features of the new \textsl{LL-branes} drastically
distinguish them from ordinary Nambu-Goto branes: 

(a) They describe intrinsically lightlike modes, whereas Nambu-Goto branes describe
massive ones.

(b) The tension of the \textsl{LL-brane} arises as an {\em additional
dynamical degree of freedom}, whereas Nambu-Goto brane tension is a given
{\em ad hoc} constant. 
The latter characteristic feature significantly distinguishes our \textsl{LL-brane}
models from the previously proposed {\em tensionless} $p$-branes (for a review,
see Ref.\ct{lindstroem-etal}). The latter rather resemble $p$-dimensional continuous
distributions of independent massless point-particles without cohesion among the latter.

(c) Consistency of \textsl{LL-brane} dynamics in a spherically or axially
symmetric gravitational background of codimension one requires the presence
of an event horizon which is automatically occupied by the \textsl{LL-brane}
(``horizon straddling'' according to the terminology of Ref.\ct{Barrabes-Israel}).

(d) When the \textsl{LL-brane} moves as a {\em test} brane in spherically or 
axially symmetric gravitational backgrounds its dynamical tension exhibits 
exponential ``inflation/deflation'' time behavior \ct{inflation-all}
-- an effect similar to the ``mass inflation'' effect around black hole horizons
\ct{israel-poisson}. 

An intriguing novel application of \textsl{LL-branes} as natural self-consistent 
gravitational sources for {\em wormhole} space-times has been developed  in a series of 
recent papers \ct{our-WH,rot-WH,ER-bridge,varna-09}.

Before proceeding let us recall that the concept of ``wormhole space-time'' 
was born in the classic work
of Einstein and Rosen \ct{einstein-rosen}, where they considered matching along
the horizon of two identical copies of the exterior Schwarzschild space-time
region (subsequently called {\em Einstein-Rosen ``bridge''}).
Another corner stone in wormhole physics is the seminal work of Morris and
Thorne \ct{morris-thorne}, who studied for the first time {\em traversable Lorentzian
wormholes}.








In what follows, when discussing wormholes we will have in mind the physically 
important class of ``thin-shell'' traversable Lorentzian wormholes first introduced by
Visser \ct{visser-thin,visser-book}.
For a comprehensive review of wormhole space-times, see Refs.\ct{visser-book,WH-rev}.

In our earlier work \ct{our-WH,rot-WH,ER-bridge,varna-09} we have constructed
various types of wormhole solutions in self-consistent systems of bulk gravity 
and bulk gauge fields (Maxwell and Kalb-Ramond) coupled to \textsl{LL-branes} 
where the latter provide the appropriate stress energy tensors, electric
currents and dynamically generated space-varying cosmological constant terms
consistently derived from well-defined world-volume \textsl{LL-brane} Lagrangian actions.

The original Einstein-Rosen ``bridge'' manifold \ct{einstein-rosen} appears as a 
particular case of the construction of spherically symmetric wormholes produced by
\textsl{LL-branes} as gravitational sources occupying the wormhole throats
(Refs.\ct{ER-bridge,rot-WH}). Thus, we are lead to the important conclusion
that consistency of Einstein equations of motion yielding the original Einstein-Rosen 
``bridge'' as well-defined solution necessarily requires the presence of 
\textsl{LL-brane} energy-momentum tensor as a source on the right hand side.

More complicated examples of spherically and axially symmetric wormholes with
Reissner-Nordstr{\"o}m and rotating cylindrical geometry, respectively,
have been explicitly constructed via \textsl{LL-branes} in Refs.\ct{our-WH,rot-WH}. 
Namely, two copies of the exterior space-time region of a Reissner-Nordstr{\"o}m or
rotating cylindrical black hole, respectively, 
are matched via \textsl{LL-brane} along what used to be the outer horizon of the
respective full black hole space-time manifold. In this way one
obtains a wormhole solution which combines the features of the Einstein-Rosen
``bridge'' on the one hand (with wormhole throat at horizon), and the features of
Misner-Wheeler wormholes \ct{misner-wheeler}, \textsl{i.e.}, exhibiting the so called 
``charge without charge'' phenomenon.

Recently the results of Refs.\ct{our-WH,rot-WH} have been  
extended to the case of {\em asymmetric} wormholes, describing two
``universes'' with different spherically symmetric geometries of black hole
type connected via a ``throat'' materialized by the pertinent gravitational
source -- an electrically charged \textsl{LL-brane}, sitting on their common
horizon. As a result of the well-defined world-volume \textsl{LL-brane} dynamics 
coupled self-consistently to gravity and bulk space-time gauge fields, it creates a
``left universe'' comprising the exterior Schwarzschild-de-Sitter space-time
region beyond the Schwarzschild horizon and where the cosmological constant is 
dynamically generated, and a ``right universe'' comprising the exterior 
Reissner-Nordstr{\"o}m region beyond the outer Reissner-Nordstr{\"o}m horizon
with dynamically generated Coulomb field-strength. Both ``universes'' are
glued together by the \textsl{LL-brane} occupying their common horizon.
Similarly, the \textsl{LL-brane} can dynamically generate a non-zero
cosmological constant in the ``right universe'', in which case it connects a
purely Schwarzschild ``left universe'' with a Reissner-Nordstr{\"o}m-de-Sitter
``right universe''.

In the present paper we will further broaden the application of \textsl{LL-branes}
in the context of wormhole physics by constructing a new type of wormhole
solution to Einstein-Maxwell equations describing a ``right universe'', which
comprises the exterior Reissner-Nordstr{\"o}m space-time region beyond the outer 
Reissner-Nordstr{\"o}m horizon, connected through a
``throat'' materialized by a \textsl{LL-brane} with a ``left universe''
being a Bertotti-Robinson space-time with two compactified spatial dimensions
\ct{BR} (see also \ct{lapedes-78}).

Let us note that previously the junction of a compactified space-time (of
Bertotti-Robinson type) to an uncompactified space-time through a wormhole
has been studied in a different setting using {\em timelike} matter on the 
junction hypersurface \ct{eduardo-GRG}. Also, in a different context a
string-like (flux tube) object with similar features to Bertotti-Robinson
solution has been constructed \ct{dzhunu} which interpolates between
uncompactified space-time regions.
\section{World-Volume Formulation of Lightlike Brane Dynamics}

There exist two equivalent {\em dual to each other} manifestly
reparametrization invariant world-volume Lagrangian formulations of 
\textsl{LL-branes} \ct{LL-brane-main,inflation-all,our-WH,rot-WH,ER-bridge,reg-BH}. 
First, let us consider the Polyakov-type
formulation where the \textsl{LL-brane} world-volume action is given as:
\be
S_{\mathrm{LL}} = \int d^{p+1}\s\,\P\llb -\h\g^{ab} g_{ab} + L\!\( F^2\)\rrb \; .
\lab{LL-action}
\ee
Here the following notions and notations are used:

(a) $\P$ is alternative non-Riemannian integration measure density (volume form) 
on the $p$-brane world-volume manifold:
\br
\P \equiv \frac{1}{(p+1)!} 
\vareps^{a_1\ldots a_{p+1}} H_{a_1\ldots a_{p+1}}(B) \; ,
\lab{mod-measure-p} \\
H_{a_1\ldots a_{p+1}}(B) = (p+1) \pa_{[a_1} B_{a_2\ldots a_{p+1}]} \; ,
\er
instead of the usual $\sqrt{-\g}$. Here $\vareps^{a_1\ldots a_{p+1}}$ is the
alternating symbol ($\vareps^{0 1\ldots p} = 1$), $\g_{ab}$ ($a,b=0,1,{\ldots},p$)
indicates the intrinsic Riemannian metric on the world-volume, and
$\g = \det\Vert\g_{ab}\Vert$.
$H_{a_1\ldots a_{p+1}}(B)$ denotes the field-strength of an auxiliary
world-volume antisymmetric tensor gauge field $B_{a_1\ldots a_{p}}$ of rank $p$.
As a special case one can build $H_{a_1\ldots a_{p+1}}$ in terms of 
$p+1$ auxiliary world-volume scalar fields $\lcurl \vp^I \rcurl_{I=1}^{p+1}$:
\be
H_{a_1\ldots a_{p+1}} = \vareps_{I_1\ldots I_{p+1}}
\pa_{a_1} \vp^{I_1}\ldots \pa_{a_{p+1}} \vp^{I_{p+1}} \;.
\lab{mod-measure-p-scalar}
\ee
Note that $\g_{ab}$ is {\em independent} of the auxiliary world-volume fields
$B_{a_1\ldots a_{p}}$ or $\vp^I$.
The alternative non-Riemannian volume form \rf{mod-measure-p}
has been first introduced in the context of modified standard (non-lightlike) string and
$p$-brane models in Refs.\ct{mod-measure}.

(b) $X^\m (\s)$ are the $p$-brane embedding coordinates in the bulk
$D$-dimensional space time with bulk Riemannian metric
$G_{\m\n}(X)$ with $\m,\n = 0,1,\ldots ,D-1$; 
$(\s)\equiv \(\s^0 \equiv \t,\s^i\)$ with $i=1,\ldots ,p$;
$\pa_a \equiv \partder{}{\s^a}$.

(c) $g_{ab}$ is the induced metric on world-volume:
\be
g_{ab} \equiv \pa_a X^{\m} \pa_b X^{\n} G_{\m\n}(X) \; ,
\lab{ind-metric}
\ee
which becomes {\em singular} on-shell (manifestation of the lightlike nature, 
cf. second Eq.\rf{on-shell-singular} below).

(d) $L\!\( F^2\)$ is the Lagrangian density of another
auxiliary $(p-1)$-rank antisymmetric tensor gauge field $A_{a_1\ldots a_{p-1}}$
on the world-volume with $p$-rank field-strength and its dual:
\be
F_{a_1 \ldots a_{p}} = p \pa_{[a_1} A_{a_2\ldots a_{p}]} \;\; ,\;\;
F^{\ast a} = \frac{1}{p!} \frac{\vareps^{a a_1\ldots a_p}}{\sqrt{-\g}}
F_{a_1 \ldots a_{p}}  \; .
\lab{p-rank}
\ee
$L\!\( F^2\)$ is {\em arbitrary} function of $F^2$ with the short-hand notation:
$F^2 \equiv F_{a_1 \ldots a_{p}} F_{b_1 \ldots b_{p}} \g^{a_1 b_1} \ldots \g^{a_p b_p}$.


Rewriting the action \rf{LL-action} in the following equivalent form:
\br
S = - \int d^{p+1}\!\s \,\chi \sqrt{-\g}
\Bigl\lb \h \g^{ab} g_{ab} - L\!\( F^2\)\Bigr\rb \; ,
\nonu \\
\chi \equiv \frac{\P}{\sqrt{-\g}} \phantom{aaaaaaaaaaaaaa}
\lab{LL-action-chi}
\er
with $\P$ the same as in \rf{mod-measure-p},
we find that the composite field $\chi$ plays the role of a {\em dynamical
(variable) brane tension}\foot{The notion of dynamical brane tension has previously 
appeared in different contexts in Refs.\ct{townsend-etal}.}.

Let us now consider the equations of motion corresponding to \rf{LL-action} 
w.r.t. $B_{a_1\ldots a_{p}}$:
\be
\pa_a \Bigl\lb \h \g^{cd} g_{cd} - L(F^2)\Bigr\rb = 0 \;\; \to \;\;
\h \g^{cd} g_{cd} - L(F^2) = M  \; ,
\lab{phi-eqs}
\ee
where $M$ is an arbitrary integration constant. The equations of motion w.r.t.
$\g^{ab}$ read:
\be
\h g_{ab} - F^2 L^{\pr}(F^2) \llb\g_{ab} 
- \frac{F^{*}_a F^{*}_b}{F^{*\, 2}}\rrb = 0  \; ,
\lab{gamma-eqs}
\ee
where $F^{*\, a}$ is the dual field strength \rf{p-rank}. Eqs.\rf{gamma-eqs} can be 
viewed as $p$-brane analogues of the string Virasoro constraints.

%
Taking the trace in \rf{gamma-eqs} and comparing with \rf{phi-eqs} 
implies the following crucial relation for the Lagrangian function $L\( F^2\)$: 
$L\!\( F^2\) - p F^2 L^\pr\!\( F^2\) + M = 0$, 
which determines $F^2$ 
on-shell as certain function of the integration
constant $M$ \rf{phi-eqs}, \textsl{i.e.} $F^2 = F^2 (M) = \mathrm{const}$.
Here and below $L^\pr(F^2)$ denotes derivative of $L(F^2)$ w.r.t. the 
argument $F^2$.

The next and most profound consequence of Eqs.\rf{gamma-eqs} is that the induced 
metric \rf{ind-metric} on the world-volume of the $p$-brane model \rf{LL-action} 
is {\em singular} on-shell (as opposed to the induced metric in the case of 
ordinary Nambu-Goto branes):
\be
g_{ab}F^{*\, b} \equiv \pa_a X^\m G_{\m\n} \(\pa_b X^\n F^{*\, b}\) =0 \; .
\lab{on-shell-singular}
\ee
Eq.\rf{on-shell-singular} is the manifestation of the {\em lightlike} nature
of the $p$-brane model \rf{LL-action} (or \rf{LL-action-chi}),
namely, the tangent vector to the world-volume $F^{*\, a}\pa_a X^\m$
is {\em lightlike} w.r.t. metric of the embedding space-time.

Further, the equations of motion w.r.t. world-volume gauge field 
$A_{a_1\ldots a_{p-1}}$ (with $\chi$ as defined in \rf{LL-action-chi} read:
\be
\pa_{[a}\( F^{\ast}_{b]}\, \chi\) = 0  \; .
\lab{A-eqs-0}
\ee

Finally, the $X^\m$ equations of motion produced by the \rf{LL-action} read:
\be
\pa_a \(\chi \sqrt{-\g} \g^{ab} \pa_b X^\m\) + 
\chi \sqrt{-\g} \g^{ab} \pa_a X^\n \pa_b X^\l \G^\m_{\n\l} = 0  \;
\lab{X-eqs-0}
\ee
where $\G^\m_{\n\l}=\h G^{\m\k}\(\pa_\n G_{\k\l}+\pa_\l G_{\k\n}-\pa_\k G_{\n\l}\)$
is the Christoffel connection for the external metric.


Eq.\rf{A-eqs-0} allows us to introduce the dual ``gauge'' potential $u$ (dual
w.r.t. world-volume gauge field $A_{a_1\ldots a_{p-1}}$ \rf{p-rank}) :
\be
F^{\ast}_{a} = c_p\, \frac{1}{\chi} \pa_a u \quad ,\quad
c_p = \mathrm{const} \; . 
\lab{u-def}
\ee
Relation \rf{u-def} enables us to rewrite Eq.\rf{gamma-eqs} (the lightlike constraint)
in terms of the dual potential $u$ in the form:
\br
\g_{ab} = \frac{1}{2a_0}\, g_{ab} - \frac{(2a_0)^{p-2}}{\chi^2}\,\pa_a u \pa_b u
\nonu \\
a_0 \equiv F^2 L^{\pr}\( F^2\)\bv_{F^2=F^2(M)} = \mathrm{const} \; .
\lab{gamma-eqs-u}
\er
($L^\pr(F^2)$ denotes derivative of $L(F^2)$ w.r.t. the argument $F^2$).
From \rf{u-def} 
we obtain the relation: 
\be
\chi^2 = -(2a_0)^{p-2} \g^{ab} \pa_a u \pa_b u \; ,
\lab{chi2-eq}
\ee
and the Bianchi identity $\nabla_a F^{\ast\, a}=0$ becomes:
\be
\pa_a \Bigl( \frac{1}{\chi}\sqrt{-\g} \g^{ab}\pa_b u\Bigr) = 0  \; .
\lab{Bianchi-id}
\ee

It is straightforward to prove that the system of equations \rf{X-eqs-0},
\rf{Bianchi-id} and \rf{chi2-eq} for $\( X^\m,u,\chi\)$, which are equivalent to the 
equations of motion \rf{phi-eqs}--\rf{A-eqs-0},\rf{X-eqs-0} resulting from the 
original Polyakov-type \textsl{LL-brane} action \rf{LL-action}, can be equivalently 
derived from the following {\em dual} Nambu-Goto-type world-volume action:
\be
S_{\rm NG} = - \int d^{p+1}\s \, T 
\sqrt{\bgv\, \det\Vert g_{ab} - \eps \frac{1}{T^2}\pa_a u \pa_b u\Vert\,\bgv} \;\; ,
\lab{LL-action-NG-A}
\ee
with $\eps = \pm 1$.
Here again $g_{ab}$ indicates the induced metric on the world-volume \rf{ind-metric}
and $T$ is dynamical variable tension simply proportional to $\chi$ 
($\chi^2 = (2 a_0)^{p-1} T^2$with $a_0$ as in \rf{gamma-eqs-u}).
The choice of the sign in \rf{LL-action-NG-A} does not have physical effect
because of the non-dynamical nature of the $u$-field.

Henceforth we will stick to the Polyakov-type formulation of world-volume
\textsl{LL-brane} dynamics since within this framework one can add in a
natural way \ct{LL-brane-main,inflation-all,our-WH}
couplings of the \textsl{LL-brane} to bulk space-time Maxwell $\cA_\m$ 
and Kalb-Ramond $\cA_{\m_1\ldots\m_{D-1}}$ gauge fields (in the case of codimension 
one \textsl{LL-branes}, \textsl{i.e.}, for $D=(p+1)+1$):
\br
{\wti S}_{\mathrm{LL}} = S_{\mathrm{LL}}
- q \int d^{p+1}\s\,\vareps^{ab_1\ldots b_p} F_{b_1\ldots b_p} \pa_a X^\m \cA_\m
\nonu \\
- \frac{\b}{(p+1)!} \int d^{p+1}\s\,\vareps^{a_1\ldots a_{p+1}}
\pa_{a_1} X^{\m_1}\ldots\pa_{a_{p+1}} X^{\m_{p+1}}
\nonu
\er
\be
\times \cA_{\m_1\ldots\m_{p+1}}
\lab{LL-action+EM+KR}
\ee
with $S_{\mathrm{LL}}$ as in \rf{LL-action}. The \textsl{LL-brane}
constraint equations \rf{phi-eqs}--\rf{gamma-eqs} are not affected by the bulk 
space-time gauge field couplings whereas Eqs.\rf{A-eqs-0}--\rf{X-eqs-0} acquire the form:
\br
\pa_{[a}\( F^{\ast}_{b]}\, \chi L^\pr (F^2)\) 
+ \frac{q}{4}\pa_a X^\m \pa_b X^\n \cF_{\m\n} = 0  \; ;
\lab{A-eqs} \\
\pa_a \(\chi \sqrt{-\g} \g^{ab} \pa_b X^\m\) + 
\chi \sqrt{-\g} \g^{ab} \pa_a X^\n \pa_b X^\l \G^\m_{\n\l}
\nonu \\
-q \vareps^{ab_1\ldots b_p} F_{b_1\ldots b_p} \pa_a X^\n \cF_{\l\n}G^{\l\m}
\nonu \\
- \frac{\b}{(p+1)!} \vareps^{a_1\ldots a_{p+1}} \pa_{a_1} X^{\m_1} \ldots
\pa_{a_{p+1}} X^{\m_{p+1}}
\nonu \\
\times \cF_{\l\m_1\dots\m_{p+1}} G^{\l\m} = 0 \; .
\lab{X-eqs}
\er
Here $\chi$ is the dynamical brane tension as in \rf{LL-action-chi},
$\cF_{\m\n} = \pa_\m \cA_\n - \pa_\n \cA_\m$ and 
\be
\cF_{\m_1\ldots\m_D} = D\pa_{[\m_1} \cA_{\m_2\ldots\m_D]} =
\cF \sqrt{-G} \vareps_{\m_1\ldots\m_D}
\lab{F-KR}
\ee
are the field-strengths of the electromagnetic $\cA_\m$ and Kalb-Ramond 
$\cA_{\m_1\ldots\m_{D-1}}$ gauge potentials \ct{aurilia-townsend}.

\section{Lightlike Brane Dynamics in Various Types of Gravitational Backgrounds}

World-volume reparametrization invariance allows us to introduce the standard 
synchronous gauge-fixing conditions:
\be
\g^{0i} = 0 \;\; (i=1,\ldots,p) \; ,\; \g^{00} = -1 \; .
\lab{gauge-fix}
\ee
Also, we will use a natural ansatz for the ``electric'' part of the 
auxiliary world-volume gauge field-strength \rf{p-rank}:
\be
F^{\ast i}= 0 \;\; (i=1,{\ldots},p) \quad ,\quad \mathrm{i.e.} \;\;
F_{0 i_1 \ldots i_{p-1}} = 0 \; ,
\lab{F-ansatz}
\ee
meaning that we choose the lightlike direction in Eq.\rf{on-shell-singular} 
to coincide with the brane
proper-time direction on the world-volume ($F^{*\, a}\pa_a \sim \pa_\t$).
The Bianchi identity ($\nabla_a F^{\ast\, a}=0$) together with 
\rf{gauge-fix}--\rf{F-ansatz} and the definition for the dual field-strength
in \rf{p-rank} imply:
\be
\pa_\t \g^{(p)} = 0 \quad \mathrm{where}\;\; \g^{(p)} \equiv \det\Vert\g_{ij}\Vert \; .
\lab{gamma-p-0}
\ee

Taking into account \rf{gauge-fix}--\rf{F-ansatz}, Eqs.\rf{gamma-eqs}
acquire the following gauge-fixed form (recall definition of the induced metric
$g_{ab}$ \rf{ind-metric}):
\be
g_{00}\equiv \Xdot^\m\!\! G_{\m\n}\!\! \Xdot^\n = 0 \quad ,\quad g_{0i} = 0 \quad ,\quad
g_{ij} - 2a_0\, \g_{ij} = 0 \; ,
\lab{gamma-eqs-0}
\ee
where $a_0$ is the same constant as in \rf{gamma-eqs-u}.

\subsection{Spherically Symmetric Backgrounds}

Here we will be interested in static spherically symmetric solutions 
of Einstein-Maxwell equations (see Eqs.\rf{Einstein-eqs}--\rf{Maxwell-eqs} below).
We will consider the following generic form of static spherically symmetric metric:
\be
ds^2 = - A(\eta) dt^2 + \frac{d\eta^2}{A(\eta)} + 
C(\eta) h_{ij}(\vec{\th}) d\th^i d\th^j \; ,
\lab{static-spherical}
\ee
or, in Eddington-Finkelstein coordinates \ct{EFM} ($dt = dv-\frac{d\eta}{A(\eta)}$) :
\be
ds^2 = - A(\eta) dv^2 + 2 dv\,d\eta + C(\eta) h_{ij}(\vec{\th}) d\th^i d\th^j \; .
\lab{EF-metric}
\ee
Here $h_{ij}$ indicates the standard metric on the sphere $S^p$. 
The radial-like coordinate $\eta$ will vary in general from $-\infty$ to $+\infty$. 

We will consider the simplest ansatz for the \textsl{LL-brane} embedding
coordinates:
\br
X^0\equiv v = \t \quad, \quad X^1\equiv \eta = \eta (\t)
\nonu \\
X^i\equiv \th^i = \s^i \;\; (i=1,\ldots ,p) \; .
\lab{X-embed}
\er
Now, the \textsl{LL-brane} equations \rf{gamma-eqs-0} together with \rf{gamma-p-0}
yield:
\be
-A(\eta) + 2\etadot = 0 \quad , \quad 
\pa_\t C = \etadot\,\pa_\eta C\bv_{\eta=\eta(\t)} = 0 \; .
\lab{eta-const}
\ee
First, we will consider the case of $C(\eta)$ as non-trivial function of $\eta$ 
(\textsl{i.e.}, proper spherically symmetric space-time). In this case 
Eqs.\rf{eta-const} imply:
\be
\etadot = 0 \; \to \; \eta (\t) = \eta_0 = \mathrm{const} \quad ,\quad 
A(\eta_0) = 0 \; .
\lab{horizon-standard}
\ee
Eq.\rf{horizon-standard} tells us that consistency of \textsl{LL-brane} dynamics in 
a proper spherically symmetric gravitational background of codimension one requires the 
latter to possess a horizon (at some $\eta = \eta_0$), which is automatically occupied 
by the \textsl{LL-brane} (``horizon straddling'' according to the
terminology of Ref.\ct{Barrabes-Israel}). Similar property -- 
``horizon straddling'', has been found also for \textsl{LL-branes} moving in
rotating axially symmetric (Kerr or Kerr-Newman) and rotating cylindrically
symmetric black hole backgrounds \ct{our-WH,rot-WH}. 

With the embedding ansatz \rf{X-embed} and assuming the bulk Maxwell field
to be purely electric static one ($\cF_{0\eta} = \cF_{v\eta} \neq 0$, the
rest being zero; this is the relevant case to be discussed in what follows), 
Eq.\rf{A-eqs} yields the simple relation:
$\pa_i \chi = 0 \;,\; \mathrm{i.e.}\;\; \chi = \chi (\t)$.
Further, the only non-trivial contribution of the second order (w.r.t. world-volume
proper time derivative) $X^\m$-equations of motion \rf{X-eqs} arises for $\m=v$, 
where the latter takes the form of an evolution equation for the dynamical tension
$\chi (\t)$. 
In the case of absence of couplings to bulk space-time gauge fields, 
the latter yields exponentional ``inflation''/``deflation'' at large times for the
dynamical \textsl{LL-brane} tension:
\be
\chi (\t) = \chi_0 
\exp\Bigl\{-\t \Bigl(\h \pa_\eta A + p a_0 \pa_\eta C \Bigr)_{\eta=\eta_0}\Bigr\}\;\; ,
\lab{chi-eq-standard-sol}
\ee
$\chi_0 = \mathrm{const}$. Similarly to the ``horizon straddling'' property, exponential 
``inflation''/``deflation'' for the \textsl{LL-brane} tension has also been
found in the case of test \textsl{LL-brane} motion in rotating axially symmetric
and rotating cylindrically symmetric black hole backgrounds 
(for details we refer to Refs.\ct{inflation-all,our-WH,rot-WH}).
This phenomenon is an analog of the ``mass inflation'' effect around black hole
horizons \ct{israel-poisson}.

\subsection{Product-Type Gravitational Backgrounds: Bertotti-Robinson Space-Time}

Consider now the case $C(\eta) = \mathrm{const}$ in \rf{EF-metric},
\textsl{i.e.}, the
corresponding space-time manifold is of product type $\S_2 \times S^p$.
A physically relevant example is the Bertotti-Robinson \ct{BR,lapedes-78}
space-time in $D=4$ (\textsl{i.e.}, $p=2$) with (non-extremal) metric 
(cf.\ct{lapedes-78}) : 
\be
ds^2 = r_0^2 \llb -\eta^2 dt^2 + \frac{d\eta^2}{\eta^2} + 
d\th^2 + \sin^2 \th d\vp^2 \rrb \; ,
\lab{BR-new}
\ee
or in Eddington-Finkelstein (EF) form ($dt = \frac{1}{r_0^2}dv - \frac{d\eta}{\eta^2}$):
\be
ds^2 = -\frac{\eta^2}{r_0^2}dv^2 + 2 dv d\eta + 
r_0^2 \llb d\th^2 + \sin^2 \th d\vp^2 \rrb \; .
\lab{BR-EF}
\ee
At $\eta = 0$ the Bertotti-Robinson metric \rf{BR-new} (or \rf{BR-EF}) possesses a 
horizon. Further, we will consider the case of Bertotti-Robinson universe with
constant electric field $\cF_{v\eta} = \pm \frac{1}{2 r_0 \sqrt{\pi}}$.
In the present case the second Eq.\rf{eta-const} is trivially satisfied whereas the
first one yields:
$\eta (\t) = \eta (0) \Bigl(1 - \t\frac{\eta (0)}{2 r_0^2}\Bigr)^{-1}$.
In particular, if the \textsl{LL-brane} is initially (at $\t=0$) located on the 
Bertotti-Robinson horizon $\eta = 0$, it will stay there permanently.

\section{Self-Consistent Wormhole Solutions Produced By Lightlike Branes}

\subsection{Lagrangian Formulation of Bulk Gravity-Matter System Coupled to 
Lightlike Brane}

Let us now consider elf-consistent bulk Einstein-Maxwell-Kalb-Ramond system coupled 
to a charged codimension-one {\em lightlike} $p$-brane (\textsl{i.e.},
$D=(p+1)+1$). It is described by the following action:
\br
S = \int\!\! d^D x\,\sqrt{-G}\,\llb \frac{R(G)}{16\pi} 
- \frac{1}{4} \cF_{\m\n}\cF^{\m\n} \right.
\nonu \\
\left. - \frac{1}{D! 2} \cF_{\m_1\ldots\m_D}\cF^{\m_1\ldots\m_D}\rrb 
+ {\wti S}_{\mathrm{LL}} \; .
\lab{E-M-KR+LL}
\er
Here $\cF_{\m\n}$ and $\cF_{\m_1\ldots\m_D}$ are the Maxwell and Kalb-Ramond
field-strengths \rf{F-KR}.
The last term on the r.h.s. of \rf{E-M-KR+LL} indicates the reparametrization
invariant world-volume action \rf{LL-action+EM+KR} 
of the \textsl{LL-brane} coupled to the bulk space-time gauge fields.

The pertinent Einstein-Maxwell-Kalb-Ramond equations of motion derived from
the action \rf{E-M-KR+LL} read:
\be
R_{\m\n} - \h G_{\m\n} R =
8\pi \( T^{(EM)}_{\m\n} + T^{(KR)}_{\m\n} + T^{(brane)}_{\m\n}\) \; ,
\lab{Einstein-eqs}
\ee
\br
\pa_\n \(\sqrt{-G}\cF^{\m\n}\) + 
q \int\!\! d^{p+1}\s\,\d^{(D)}\Bigl(x-X(\s)\Bigr)
\nonu \\
\times\; \vareps^{ab_1\ldots b_p} F_{b_1\ldots b_p} \pa_a X^\m = 0 \; ,
\lab{Maxwell-eqs} \\
\vareps^{\n\m_1\ldots\m_{p+1}} \pa_\n \cF - 
\b\,\int\! d^{p+1}\s\,\d^{(D)}(x - X(\s))
\nonu \\
\times\; \vareps^{a_1\ldots a_{p+1}} \pa_{a_1}X^{\m_1}\ldots\pa_{a_{p+1}}X^{\m_{p+1}} = 0 \; ,
\lab{F-KR-eqs}
\er
where in the last equation we have used the last relation \rf{F-KR}. 
The explicit form of the energy-momentum tensors read:
\br
T^{(EM)}_{\m\n} = \cF_{\m\k}\cF_{\n\l} G^{\k\l} - G_{\m\n}\frac{1}{4}
\cF_{\r\k}\cF_{\s\l} G^{\r\s}G^{\k\l} \; ,
\lab{T-EM} \\
T^{(KR)}_{\m\n} = \frac{1}{(D-1)!}\llb \cF_{\m\l_1\ldots\l_{D-1}}
{\cF_{\n}}^{\l_1\ldots\l_{D-1}} \right.
\nonu \\
\left. - \frac{1}{2D} G_{\m\n} \cF_{\l_1\ldots\l_D} \cF^{\l_1\ldots\l_D}\rrb
= - \h \cF^2 G_{\m\n}  \; ,
\lab{T-KR} \\
T^{(brane)}_{\m\n} = - G_{\m\k}G_{\n\l}
\int\!\! d^{p+1}\s\,\frac{\d^{(D)}\Bigl(x-X(\s)\Bigr)}{\sqrt{-G}}
\nonu \\
\times\; \chi\,\sqrt{-\g} \g^{ab}\pa_a X^\k \pa_b X^\l \; ,
\lab{T-brane}
\er
where the brane stress-energy tensor is straightforwardly derived
from the world-volume action \rf{LL-action} (or, equivalently, \rf{LL-action-chi};
recall $\chi\equiv\frac{\P}{\sqrt{-\g}}$ is the variable brane tension).

Using again the embedding ansatz \rf{X-embed} together with \rf{horizon-standard}
as well as \rf{gauge-fix}--\rf{gamma-eqs-0}, the Kalb-Ramond equations of motion
\rf{F-KR-eqs} reduce to:
\be
\pa_\eta \cF + \b \d (\eta-\eta_0) = 0
\lab{F-KR-0}
\ee
implying
\br
\cF = \cF_{(+)} \th (\eta-\eta_0) + \cF_{(-)} \th (\eta_0 -\eta)
\nonu \\
\cF_{(\pm)} = \mathrm{const} \quad ,\quad \cF_{(-)} - \cF_{(+)} = \b
\lab{F-jump}
\er
Therefore, a space-time varying non-negative cosmological constant is dynamically 
generated in both exterior and interior regions w.r.t. the horizon at $\eta=\eta_0$
(cf. Eq.\rf{T-KR}): $\L_{(\pm)} = 4\pi \cF^2_{(\pm)}$.
Hereafter we will discard the presence of the Kalb-Ramond gauge field and,
correspondingly, there will be no dynamical generation of cosmological
constant.

\subsection{Asymmetric Wormholes}
We will consider in what follows the case of 
$D=4$-dimensional bulk space-time and, correspondingly, $p=2$ for the 
\textsl{LL-brane}. 
For further simplification of the numerical constant factors we will choose
the following specific (``wrong-sign'' Maxwell) form for the Lagrangian of the 
auxiliary non-dynamical world-volume gauge field \rf{p-rank}:
$L(F^2)=\frac{1}{4}F^2 \;\; \to \;\;  a_0 = M$,
where again $a_0$ is the constant defined in \rf{gamma-eqs-u} and
$M$ denotes the original integration constant in Eqs.\rf{phi-eqs}.

We will seek a self-consistent solution of the equations of motion of the
coupled Einstein-Maxwell-\textsl{LL-brane} system 
(Eqs.\rf{Einstein-eqs}--\rf{Maxwell-eqs} and \rf{phi-eqs}--\rf{gamma-eqs}, 
\rf{A-eqs}--\rf{X-eqs}) describing an asymmetric wormhole space-time with spherically
symmetric geometry. The general form of asymmetric wormhole metric (in
Eddington-Finkelstein coordinates) reads:
\be
ds^2 = - A(\eta) dv^2 + 2 dv d\eta + 
C(\eta) \llb d\th^2 + \sin^2 \th d\vp^2 \rrb \, ,
\lab{asymm-WH-EF}
\ee
\br
A (0) = 0 \;\;\; (\; \mathrm{``throat'' ~at}\; \eta_0 =0 \; )
\nonu \\
A(\eta) > 0 \;\; \mathrm{for ~all}\; \eta \neq 0 \; .
\lab{asymm-WH-A}
\er
The radial-like coordinate $\eta$ varies from $-\infty$ to $+\infty$ and the
metric coefficients $A(\eta)$ and $C(\eta)$ are continuous but 
{\em not necessarily differentiable} w.r.t. $\eta$ at the wormhole ``throat''
$\eta = 0$. We will require:
\be
\pa_\eta A\bv_{\eta \to +0} \equiv \pa_\eta A \bv_{+0} >0 \; , \; 
\pa_\eta A\bv_{\eta \to -0} \equiv \pa_\eta A \bv_{-0} >0 \; .
\lab{A-der}
\ee

Einstein equations \rf{Einstein-eqs} yield for the metric \rf{asymm-WH-EF}:
\br
\pa_\eta A \bv_{+0} - \pa_\eta A \bv_{-0} = - 16\pi\,\chi
\nonu \\
\pa_\eta \ln C\bv_{+0} - \pa_\eta \ln C\bv_{-0} = - \frac{4\pi\,\chi}{a_0} \; .
\lab{Einstein-eqs-0}
\er

For the \textsl{LL-brane} equations of motion we use again the embedding \rf{X-embed}
resulting in the \textsl{LL-brane} ``horizon straddling'' \rf{horizon-standard}.
On the other hand, the second order Eqs.\rf{X-eqs} contain ``force''
terms (the geodesic ones involving the Christoffel connection coefficients as well as 
those coming from the \textsl{LL-brane} coupling to the bulk Maxwell 
gauge field) which display discontinuities across the ``throat'' at $\eta=0$
occupied by the \textsl{LL-brane} due to the delta-function terms in the respective
bulk space-time Einstein-Maxwell Eqs.\rf{Einstein-eqs}--\rf{Maxwell-eqs} 
(now $\eta_0 \equiv 0$). The discontinuity problem is resolved following 
the approach in Ref.\ct{Israel-66} (see also the regularization
approach in Ref.\ct{BGG}, Appendix A) by taking mean values of the ``force''
terms across the discontinuity at $\eta=0$. Furthermore, we will require
$\chi = \mathrm{const}$ (independent of the \textsl{LL-brane} proper time $\t$) 
for consistency with the matching relations \rf{Einstein-eqs-0}. Thus, 
in the case of the \textsl{LL-brane} embedding \rf{X-embed}
the $X^\m$-equation \rf{X-eqs} for $\m=v$ with $D=4\, ,\, p=2$, 
no Kalb-Ramond coupling, \textsl{i.e.}, $\cF = 0$, and using the gauge-fixing
\rf{gauge-fix}, becomes:
\br
\chi \Bigl\lb \frac{1}{4}\(\pa_\eta A \bv_{+0} + 
\pa_\eta A\bv_{-0}\) + a_0 \(\pa_\eta \ln C\bv_{+0} + \right.
\nonu \\
\left. + \pa_\eta \ln C\bv_{-0}\)\Bigr\rb 
- q \sqrt{2a_0} \llb \cF_{v\eta}\bv_{+0} + \cF_{v\eta}\bv_{-0}\rrb = 0
\lab{X0-eq}
\er

In the present wormhole solution we will take ``left'' Bertotti-Robinson ``universe''
with:
\be
A(\eta) = \frac{\eta^2}{r_0^2} \quad, \quad C(\eta) = r_0^2 
\quad, \quad \cF_{v\eta} = \pm \frac{1}{2\sqrt{\pi}\,r_0}
\lab{left-BR}
\ee
for $\eta <0$, and ``right'' Reissner-Nordstr{\"o}m ``universe'' with:
\br
A(\eta) \equiv A_{\mathrm{RN}}(r_0 + \eta) = 
1 - \frac{2m}{r_0 + \eta} + \frac{Q^2}{(r_0 + \eta)^2} \; ,
\nonu
\er
\be
C(\eta) = (r_0 + \eta)^2 \; ,\;
\cF_{v\eta} \equiv \cF_{vr}\!\!\bv_{RN} = \frac{Q}{\sqrt{4\pi} (r_0 + \eta)^2}\; ,
\lab{right-RN-1}
\ee
for $\eta >0$, and
\br
A(0) \equiv A_{\mathrm{RN}}(r_0) = 0 \; ,\; 
\pa_\eta A \bv_{+0} \equiv \pa_r A_{\mathrm{RN}}\bv_{r=r_0} >0 
\lab{right-RN-2}
\er
where $\cF_{v\eta}$'s are the respective Maxwell field-strengths and where 
$Q = r_0 \Bigl\lb\sqrt{\frac{8\pi}{a_0}}\, q r_0 \pm 1\Bigr\rb$
is determined from the discontinuity of $\cF_{v\eta}$ in Maxwell equations
\rf{Maxwell-eqs} across the charged \textsl{LL-brane}. Here we have used the standard
coordinate notations for the Reissner-Nordstr{\"o}m metric coefficients and
Coulomb field strength:
\be
A_{\mathrm{RN}}(r) = 1 - \frac{2m}{r} + \frac{Q^2}{r^2} \quad,\quad 
\cF_{vr}\bv_{RN} = \frac{Q}{\sqrt{4\pi} r^2} \; .
\lab{RN-standard}
\ee

Since obviously both Bertotti-Robinson \rf{left-BR} and
Reissner-Nordstr{\"o}m \rf{right-RN-1} metrics do satisfy the ``vacuum''
Einstein-Maxwell equations (Eqs.\rf{Einstein-eqs}--\rf{Maxwell-eqs} {\em without} 
the \textsl{LL-brane}
stress-energy tensor) it remains to check the matching of both metrics at
the ``throat'' $\eta = 0$ (the location of the \textsl{LL-brane}) according
to Eqs.\rf{Einstein-eqs-0}--\rf{X0-eq}. In this case the latter equations give:
\br
\pa_r A_{\mathrm{RN}}\bv_{r=r_0} = - 16\pi\,\chi \;\; ,\;\;
\pa_r \ln r^2\bv_{r=r_0} = - \frac{4\pi}{a_0}\chi
\lab{Einstein-eqs-1} \\
\chi \Bigl\lb \frac{1}{4}\pa_r A_{\mathrm{RN}}\bv_{r=r_0} + 
a_0\pa_r\ln r^2\bv_{r=r_0}\Bigr\rb
\nonu \\
- 2q^2 \mp \frac{q}{r_0}\sqrt{\frac{2a_0}{\pi}} = 0 \; .
\lab{X0-eq-1}
\er
From \rf{Einstein-eqs-1}--\rf{X0-eq-1} we get:
\be
r_0 = \frac{a_0}{2\pi |\chi|} \quad ,\quad
m = \frac{a_0}{2\pi |\chi|} \( 1 - 4a_0\) \; ,
\lab{param-1}
\ee
implying that the dynamical \textsl{LL-brane} tension $\chi$ must be negative, thus
identifying the \textsl{LL-brane} as ``exotic matter'' \ct{morris-thorne,visser-book}.
Further we obtain a quadratic equation for $|\chi|$:
\be
\chi^2 + \frac{q^2}{4\pi} \pm \frac{q}{2\sqrt{2\pi\,a_0}}|\chi| = 0 \; ,
\lab{chi-eq-2}
\ee
which dictates that we have to choose the sign of $q$ to be opposite to the sign in the
expression for the Bertotti-Robinson constant electric field (last Eq.\rf{left-BR}). 
There are two positive solutions for $|\chi|$:
\be
|\chi| = \frac{|q|}{4\sqrt{2\pi\,a_0}} \( 1 \pm \sqrt{1-8a_0}\)  \quad 
\mathrm{for}\;\; a_0 < 1/8 \; .
\lab{param-2}
\ee
Using \rf{param-1} and \rf{param-2} the expression 
for $Q^2$ reads:
\be
Q^2 = \frac{a_0^2}{4\pi^2 \chi^2} \( 1 - 8a_0\) = 
\frac{8a_0^3}{\pi\, q^2}\, \frac{1 - 8a_0}{\( 1 \pm \sqrt{1 - 8a_0}\)^2}
\lab{param-3}
\ee

Thus, we have constructed a solution to Einstein-Maxwell equations
\rf{Einstein-eqs}--\rf{Maxwell-eqs} in $D=4$ describing a wormhole space-time manifold 
consisting of a ``left'' Bertotti-Robinson universe with two compactified space
dimensions and a ``right'' Reissner-Nordstr{\"o}m universe connected by a
``throat'' materialized by a \textsl{LL-brane}. The ``throat'' is a
common horizon for both universes where for the ``right'' universe it is the
external Reissner-Nordstr{\"o}m horizon.
All wormhole parameters, including the dynamical \textsl{LL-brane} tension, 
are determined in terms of the surface charge density $q$ of the \textsl{LL-brane} 
(cf. Eq.\rf{LL-action+EM+KR}) and the integration constant $a_0$ \rf{gamma-eqs-u}
characterizing \textsl{LL-brane} dynamics in a bulk gravitational field.

\section{Conclusions. Travel to Compactland Through a Wormhole}

In this work we have explored the use of (codimension-one) \textsl{LL-branes} for 
construction of new asymmetric wormhole solutions of Einstein-Maxwell equations. 
We have put strong emphasize on the crucial properties of the dynamics of 
\textsl{LL-branes} interacting with gravity and bulk space-time gauge fields:

(i) ``Horizon straddling'' -- automatic position of the \textsl{LL-brane} on (one of) the
horizon(s) of the bulk space-time geometry; 

(ii) Intrinsic nature of the \textsl{LL-brane} tension as an additional 
{\em dynamical degree of freedom} unlike the case of standard Nambu-Goto $p$-branes;

(iii) The \textsl{LL-brane} stress-energy tensor is systematically derived
from the underlying \textsl{LL-brane} Lagrangian action and provides the appropriate
source term on the r.h.s. of Einstein equations to enable the existence of
consistent non-trivial wormhole solutions; 

(iv) Electrically charged \textsl{LL-branes} naturally produce {\em asymmetric}
wormholes with the \textsl{LL-brane} itself materializing the wormhole ``throat''
and uniquely determining the pertinent wormhole parameters.



Finally, let us point out that the above asymmetric wormhole connecting
Reissner-Nordstr{\"o}m universe with a Bertotti-Robinson universe through a 
lightlike hypersurface occupied by a \textsl{LL-brane} is {\em traversable}
w.r.t. the {\em proper time} of a traveling observer. The latter property is
similar to the {\em proper time} traversability of other symmetric and
asymmetric wormholes with \textsl{LL-brane} sitting on the ``throat'' 
\ct{our-WH,rot-WH,ER-bridge,varna-09}.
Indeed, let us consider test particle (``traveling observer'') dynamics
in the asymmetric wormhole background given by \rf{left-BR}--\rf{right-RN-1}, which is
described by the action:
\be
S_{\mathrm{particle}} = \h \int d\l \Bigl\lb\frac{1}{e}\xdot^\m \xdot^\n G_{\m\n}
- e m_0^2 \rb \; .
\lab{test-particle}
\ee
Using energy $\cE$ and orbital momentum $\cJ$ conservation and introducing the 
{\em proper} world-line time $s$ ($\frac{ds}{d\l}= e m_0$), the ``mass-shell'' equation
(the equation w.r.t. the ``einbein'' $e$ produced by the action \rf{test-particle})
yields:
\be
\(\frac{d\eta}{ds}\)^2 + \cV_{\mathrm{eff}} (\eta) = \frac{\cE^2}{m_0^2}
\; ,\; 
\cV_{\mathrm{eff}} (\eta) \equiv A(\eta) \Bigl( 1 + \frac{\cJ^2}{m_0^2 C(\eta)}\Bigr) 
\lab{particle-eq-2}
\ee
with $\cA (\eta),\, C(\eta)$ -- the same metric coefficients as  in 
\rf{left-BR}--\rf{right-RN-2}.

For generic values of the parameters the effective potential in the Bertotti-Robinson
universe \rf{left-BR} (\textsl{i.e.}, for $\eta <0$) 
has harmonic-oscillator-type form. Therefore, a traveling observer 
starting in the Reissner-Nordstr{\"o}m universe \rf{right-RN-1} (\textsl{i.e.},
at some $\eta >0$) and moving ``radially'' along the $\eta$-direction towards the
horizon, will cross the wormhole ``throat''
($\eta=0$) within finite interval of his/her proper time, then will continue into the 
Bertotti-Robinson universe subject to harmonic-oscillator deceleration force, will
reverse back at the turning point and finally will cross the ``throat'' back into the
Reissner-Nordstr{\"o}m universe.







Let us stress that, as in the case of the previously constructed symmetric and
asymmetric wormholes via \textsl{LL-branes} sitting on their ``throats'' 
\ct{our-WH,rot-WH,ER-bridge,varna-09}, the present 
Reissner-Nordstr{\"o}m-to-Bertotti-Robinson wormhole is
{\em not} traversable w.r.t. the ``laboratory'' time of a static observer in
either universe.

\section*{Acknowledgments}
E.N. and S.P. are supported by Bulgarian NSF grant \textsl{DO 02-257}.
E.G. thanks the astrophysics and cosmology group at PUCV, Chile,  for hospitality.
Also, all of us acknowledge support of our collaboration through the exchange
agreement between the Ben-Gurion University of the Negev and the Bulgarian Academy 
of Sciences.


\end{multicols}
\end{document}